# Determination of Negative Permittivity and Permeability of Metamaterials from Reflection and Transmission Coefficients

D. R. Smith, S. Schultz
*Department of Physics, University of California, San Diego, 9500 Gilman Drive, La Jolla, CA 92093-0319*

P. Markoš[*], C. M. Soukoulis
*Ames Laboratory and Department of Physics and Astronomy, Iowa State University, Ames, Iowa 50011*
(Submitted 26 November 2001)

We analyze the reflection and transmission coefficients calculated from transfer matrix simulations on finite lengths of electromagnetic metamaterials, to determine the effective permittivity ($\varepsilon$) and permeability ($\mu$). We perform this analysis on structures composed of periodic arrangements of wires, split ring resonators (SRRs) and both wires and SRRs. We find the recovered frequency-dependent $\varepsilon$ and $\mu$ are entirely consistent with analytic expressions predicted by effective medium arguments. Of particular relevance are that a wire medium exhibits a frequency region in which the real part of $\varepsilon$ is negative, and SRRs produce a frequency region in which the real part of $\mu$ is negative. In the combination structure, at frequencies where both the recovered real parts of $\varepsilon$ and $\mu$ are simultaneously negative, the real part of the index-of-refraction is found also to be unambiguously negative.

It has been proposed that electromagnetic metamaterials—composite structured materials, formed from either periodic or random arrays of scattering elements—should respond to electromagnetic radiation as continuous materials, at least in the long wavelength limit [1,2]. In recent experiments and simulations [3,4], it has been demonstrated that certain metamaterial configurations exhibit scattering behavior consistent with the assumption of approximate frequency-dependent forms for $\varepsilon$ and $\mu$. However, the techniques applied in those studies probed the materials indirectly, and did not provide an explicit measurement that would assign values for $\varepsilon$ and $\mu$. It is our aim in this paper to show that the previous conjectures were indeed valid: unambiguous values for $\varepsilon$ and $\mu$ can be applied to electromagnetic metamaterials. Our approach here utilizes the transmission and reflection coefficients (S-parameters, equivalently) calculated for a wave normally incident on a finite slab of metamaterial. We invert the scattering data to determine n and z, for systems of several thicknesses, from which we obtain self-consistent values for $\varepsilon$ and $\mu$. While we utilize simulation data in this study, the technique we describe will be readily applicable to the experimental characterization of metamaterial samples whenever the scattering parameters are known.

The common method of characterizing the electromagnetic scattering properties of a homogeneous material is to identify its impedance (z) and refractive index (n) [5,6]. While it is possible to completely specify the scattering in terms of z and n, it is often more convenient to choose a second set of analytic variables that carry a direct material interpretation. These variables are the electric permittivity $\varepsilon = n/z$, and the magnetic permeability $\mu = nz$. Both n and z, and therefore $\varepsilon$ and $\mu$, are frequency-dependent complex functions that satisfy certain requirements based on causality. For passive materials, Re(z) and Im(n) must be greater than zero.

At a given frequency, any material that supports only one propagating mode will generally exhibit a well-defined refractive index n, whether the material is continuous or not. Photonic band gap materials, for example, are characterized by dispersion curves from which an effective index can extracted, even for bands well above the first band gap, and well-defined refraction phenomena can result [7,8,9]. However, it is generally not possible to assign an impedance (z) to a non-continuous material, except in those cases where the wavelength in the material is much larger than the dimensions and spacing of the constituent scattering elements that compose the medium. If z depends strongly on the surface termination, or if z depends on the overall size of the sample, then z is ambiguous, and it is not possible to assign intrinsic values for $\varepsilon$ and $\mu$.

The transmission coefficient for waves incident normally to the face of a 1-D slab of continuous material (in vacuum) with length d is related to n and z by

$$t^{-1} = \left[\cos(nkd) - \frac{i}{2}\left(z + \frac{1}{z}\right)\sin(nkd)\right]e^{ikd}, \quad (1)$$

where $k=\omega/c$ is the wavenumber of the incident wave. The incident wave is assumed to travel rightward along the positive x-axis, with the origin defined as the first face of the material seen by the radiation. To improve the clarity of the subsequent formulas, we introduce the normalized transmission coefficient $t'=\exp(ikd)t$. The reflection coefficient is also related to z and n by

$$\frac{r}{t'} = -\frac{1}{2}i\left(z - \frac{1}{z}\right)\sin(nkd) \quad (2)$$

---

[*] Permanent address: Institute of Physics, Slovak Academy of Sciences, Dúbravska cesta 9, 842 28 Bratislava, Slovakia.

Eqs. (1) and (2) can be inverted to find n and z as functions of t' and r. Performing this inversion leads to the following expressions:

$$\cos(nkd) = \frac{1}{2t'}\left[1-\left(r^2 - t'^2\right)\right]$$
$$= \text{Re}\left(\frac{1}{t'}\right) - \frac{1}{2|t'|^2}\left(A_1 r + A_2 t'\right) \quad (3)$$

and

$$z = \pm\sqrt{\frac{(1+r)^2 - t'^2}{(1-r)^2 - t'^2}}, \quad (4)$$

where $A_1 = r^*t' + t'^*r$ and $A_2 = 1 - |r|^2 - |t'|^2$ are both real-valued functions that go to zero in the absence of material losses.

Note that while the expressions for n and z are relatively uncomplicated, they are complex functions with multiple branches, the interpretation of which can lead to ambiguities in determining the final expressions for ε and μ. We can resolve these ambiguities by making use of additional knowledge about the material. For example, if the material is passive, the requirement that Re(z)>0 fixes the choice of sign in Eq. 4. Likewise, Im(n)>0 leads to an unambiguous result for Im(n):

$$\text{Im}(n) = \pm\text{Im}\left(\frac{\cos^{-1}\left(\frac{1}{2t'}\left[1-\left(r^2-t'^2\right)\right]\right)}{kd}\right) \quad (5)$$

When we solve for the right-hand side of Eq. 5, we select whichever of the two roots yields a positive solution for Im(n). Re(n), however, is complicated by the branches of the arccosine function, or,

$$\text{Re}(n) = \pm\text{Re}\left(\frac{\cos^{-1}\left(\frac{1}{2t'}\left[1-\left(r^2-t'^2\right)\right]\right)}{kd}\right) + \frac{2\pi m}{kd}, \quad (6)$$

where m is an integer. When d is large, these branches can lie arbitrarily close to one another, making the selection of the correct branch difficult in the case of dispersive materials. For this reason best results are obtained for the smallest possible thickness of sample, as has commonly been known in the analysis of continuous materials. Even with a small sample, more than one thickness must be measured to identify the correct branch(es) of the solution which yields consistently the same values for n. Note that the requirement that Im(n)>0 uniquely identifies the sign of Re(n), which is essential when the material may potentially have regions which are *left-handed*—that is, materials in which Re(n) may be negative.

In order to demonstrate the validity of assigning bulk permeability and permittivity values to non continuous metamaterials, we utilize simulation data generated by the transfer matrix method (TMM) [10,11] on three types of structures: a medium composed of a periodic array of wires, a medium composed of a periodic array of split ring resonators (SRRs), and a medium composed by interleaving the wire and the SRR arrays. We simulate one unit cell of a given system using periodic boundary conditions in the directions perpendicular to the propagation direction. The TMM represents a complete solution of Maxwell's equations, and differs from Eqs. (1) and (2) in that the exact details of the scattering elements are taken into account. In the embedding medium, all eigenvalues of the TM can be found analytically [12]. We find that in the frequency range of interest and for the incoming wave perpendicular to the sample surface, there is only one propagating mode. All other modes are evanescent, so that the systems considered are effectively one-dimensional and we are justified utilizing Eqs. (5) and (6), in which t' and r represent the transmission and reflection for this propagating mode. In the calculations presented here, cells were meshed with 15x11x15 grid points, in a manner described in Ref. 11. In the formulation of the TMM used here, the metal dielectric function is approximated by ε=-1000 + 18,000i, for all frequencies. A larger value of the imaginary part of the dielectric function for the metal results in lowered absorption, but does not change the qualitative picture presented here.

A square array of conducting wires, based on effective medium arguments, is expected to exhibit the ideal frequency-dependent plasmonic form of $\varepsilon=1-\omega_{co}^2/\omega^2$, where $\omega_{co}$ is a cutoff frequency [1]. Fig. 1 shows the permittivity recovered from simulation data for an array of wires, continuous along the E-polarized direction of the incident wave, and with lattice constant 5 mm. The figure shows data for one, three, and five layers of lattice in the propagation direction. A fit to the resulting curves indicates a cutoff frequency of $\omega_p/2\pi=22.3$ GHz, consistent with that determined by the noting the onset of transmission (not shown), with a frequency dependence indistinguishable from the three and five row curves. The recovered permeability was found to be unity at all frequencies.

While there is no ambiguity in determining the phase of the transmitted wave, the phase of the reflected wave must be measured relative to a reference plane that is considered to be at the first face of the material. Given that metamaterials generally do not have well-defined surfaces, the question arises as to where this reference plane should be located. A definite answer can be obtained in the case of the wire structure, where it is reasonable to assume that the reference plane coincides with the first plane of the wires. We found that with this choice of reference plane, the relation $z(\omega)=1/n(\omega)$ holds, consistent with the response of the wire medium being entirely electric (i.e., $\mu=1$). For the SRR medium, the choice of reference plane was not found to be critical; for the combined structure, however, the wires dominate the scattering, and the reference plane was again chosen to coincide with the first plane of wires.

A second ambiguity that arises is the effective unit cell length that will result in values of $\varepsilon$ and $\mu$ that do not depend on the overall length of the structure. Empirically, we find in the three cases investigated that the correct length corresponds to the number of unit cells multiplied by the length of each unit cell. This can be seen in Fig. 1, where the analysis of different lengths of the wire medium results in the same permittivity; only for a wire medium consisting of one layer is there a noticeable deviation from the "bulk" behavior. It should be noted that the unit cells analyzed here are square, and thus the unit cell length is implicit in the calculation by virtue of the spacing between elements in the direction perpendicular to propagation. The fact that even one unit cell in the propagation direction produces nearly the same scattering parameters as a thicker material may therefore not be so unexpected.

Fig. 2 shows the real and imaginary parts of the permeability for a medium composed of only SRRs, as configured in Ref. [11]. The dark solid curves are taken from the simulation data, while the dashed curves are from transmission and reflection coefficients computed for a uniform material with a permeability corresponding to

$$\mu(\omega) = 1 - \frac{\omega_{mp}^2}{\omega^2 - \omega_0^2 + i\Gamma\omega} \quad (7)$$

where we have used $\omega_0/2\pi=8.5$ GHz, $\omega_{mp}/2\pi=3.3$ GHz and Γ=2.0 GHz. The recovered form for the permeability is in agreement with the ideal resonant form shown in Eq. 7, indicating a region of negative permeability over the range from ~8.5 GHz to 9 GHz. This is consistent with the region of attenuated transmission found from a plot of the transmittance versus frequency (not shown).

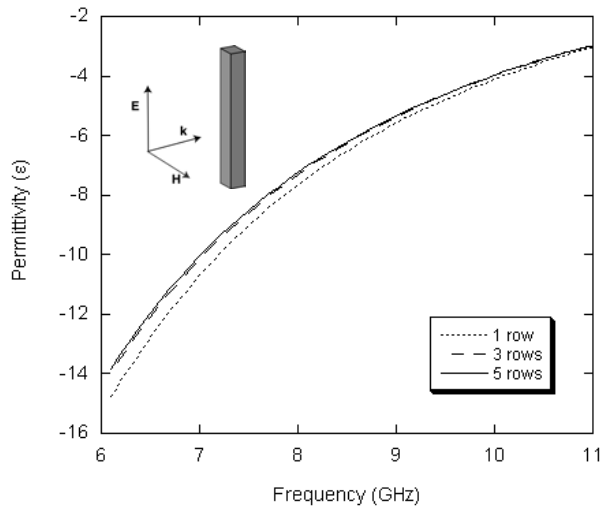

**Figure 1:** Permittivity as a function of frequency for a wire medium. The cutoff frequency for the structure, determined by the onset of propagation is 22.3 GHz. The inset depicts the orientation of the wire with respect to the incident radiation.

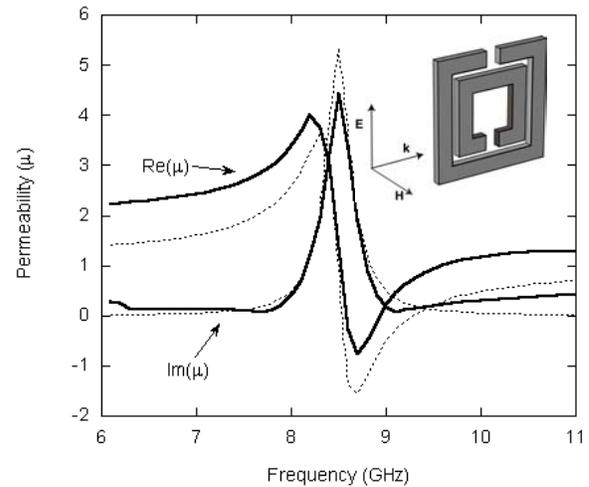

**Figure 2:** Permeability as a function of frequency for the SRR medium, for one or more unit cell lengths. The solid dark lines correspond to the real and imaginary parts of the permeability determined using the TMM data, while the dashed lines were obtained from Eq. (7) by assuming parameters chosen to approximate the SRR medium. Note that at ~8.5 GHz the real part of the permeability goes negative, with a bandwidth of ~0.5 GHz.

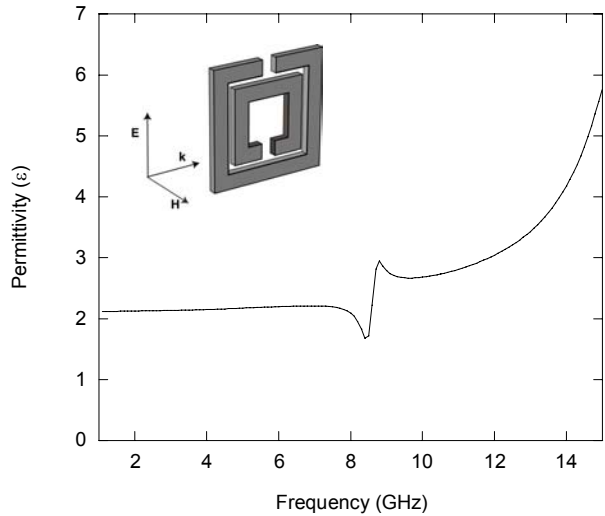

**Figure 3:** Real part of the Recovered permittivity as a function of frequency for the SRR medium.

An array of metal scattering elements can be expected to have a net polarization in response to an applied electric field, and it might be further anticipated that the effective permittivity will be dispersive, especially if the elements are resonant. In Fig. 3, we plot the recovered frequency-dependent permittivity corresponding to the SRR medium, which is significant over the frequency range covered.

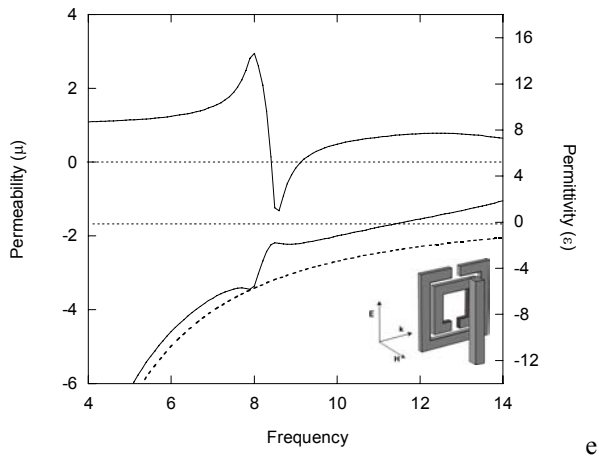

**Figure 4:** Recovered Re(μ) (top, left-hand scale) and Re(ε) (bottom, solid curve, right-hand scale) as a function of frequency for the combination SRR/wire medium. The flat dashed lines indicate the zero values for the two axes (offset for clarity). Note that while the addition of the SRR medium to the wire medium results in a more positive permittivity (solid line) than wires alone (dashed curve), nevertheless there is a frequency region (~8.5-9.0 GHz) where both ε and μ are negative.

The result of combining a wire medium with an SRR medium is shown in Fig. 4, in which we compare the real parts of the recovered permittivity and permeability from the combined structure with that from either the wire or the SRR medium alone. The solid curves are from the combined structure, while the dashed curves are from either medium alone. Consistent with what is observed in other calculations, as well as experiments, the addition of the wires to the SRR medium does not impact the permeability properties of the SRR medium. However, the combination of the SRRs and wires results in a permittivity that is less negative than that for wires alone.

It was first theoretically suggested by Veselago [13] that an incoming wave incident on an interface between vacuum and material with simultaneously negative Re(ε) and Re(μ) would be refracted to the same side of the normal to the interface. By considering the analytic properties of the refractive index in such a medium, it was shown that the refractive index n must, in fact, be negative wherever ε and μ are both negative [14]. In a recent experiment [4], a metamaterial formed by combining a wire medium with a SRR medium, qualitatively similar to those simulated here, was used to demonstrate negative refraction, and the angle of refraction was utilized to recover n as a function of frequency. While a refraction experiment is a useful method to measure the index, we can also obtain this information directly and unambiguously using Eq. 3, as shown in Fig. 5. If one has access to phase information in the transmission and reflection coefficients, one can identify both the magnitude *and sign* of the refractive index from direct incidence measurements on a planar slab.

We have demonstrated that the traditional procedure of obtaining material parameters from transmission/reflection data can be successfully applied to metamaterials, subject to the possible ambiguities of the definition of the first surface of the measured sample, and the sample length. The former ambiguity can be resolved in certain cases by using additional information about the material, as shown in the case for the wire medium above. In cases for which it is not possible to determine the correct position of the scattering surface, the range of possible solutions corresponding to the various possible surfaces throughout the unit cell will place bounds on the possible values of the effective material parameters. These bounds will become narrower as the unit cell becomes smaller in comparison with the wavelength. The correct length of a unit cell can be determined

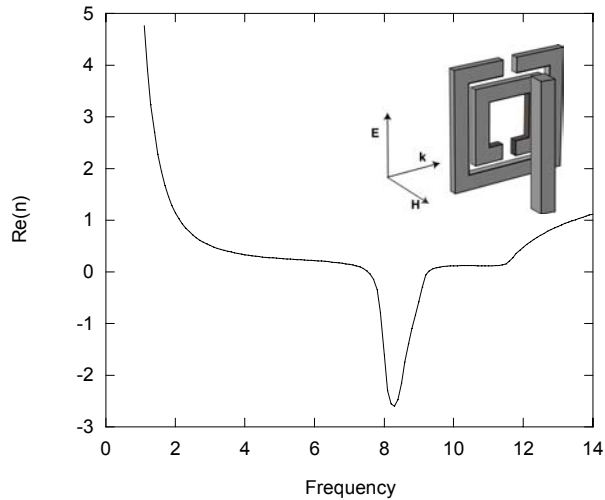

**Figure 5:** Refractive index (n) as a function of frequency for the combination SRR/wire medium. The region where the index is negative corresponds to the frequency region where both ε and μ are simultaneously negative.

empirically, by requiring a solution that is stable for several analyzed lengths.

As a final point, we note that this analysis depends entirely on our ability to measure structures for which the electric polarization **P** depends only on the electric field **E**, and the magnetization **M** depends only on the magnetic field **H**. For a general scattering element, this does not necessarily hold, and we may have, for example, a constitutive relation such as **H**=γ**E**+**B**/μ [15]. A medium that responds in such a manner will exhibit chirality, and a rotation of the polarization. Indeed, in these calculations we do find some degree of polarization conversion in the data for the SRRs and the combined SRR/wire medium. This evidence of chiral behavior suggests the analysis presented here provides a good, but not exact, characterization of the metamaterial. Further improvements to the technique should allow a full characterization to be implemented.

ACKNOWLEDGEMENTS:
We are grateful for helpful discussions with Art Vetter (Boeing, PhantomWorks), and John Pendry (Imperial College). This work was supported by DARPA, through grants from AFOSR (Contract No. KG3523) and ARO (Contract No. DAAD19-00-1-0525). CS also thanks NATO and Ames Lab. (DOE Contract No. W-7405-Eng-82) for support, and PM thanks the Slovak Grant Agency for partial financial support.